\documentstyle[11pt]{article}
\def\th{\theta} 

\def\k{\kappa}
\def\a{\alpha}
\def\b{\beta}
\def\d{\delta} 

\def\ve{\varepsilon}

\def\k{\kappa}
\def\l{\lambda} 
\def\m{\mu}

\newcommand{\cN}{\mathcal{N}}
\newcommand{\cO}{{\cal O}}
\def\be{\begin{equation}}
\def\ee{\end{equation}}
\def\td{\tilde}
\def\bea{\begin{eqnarray}}
\def\eea{\end{eqnarray}}
\def\loon{\relbar\joinrel\relbar\joinrel\rightarrow}

\begin{document}

\begin{flushright} BRX TH-583 \end{flushright}


\begin{center}
{\Large\bf Reggeization of  $\cN$=8 Supergravity and \\
 $\cN$=4 Yang--Mills Theory}

{\large Howard J.\ Schnitzer}\footnote{email:
schnitzr@brandeis.edu\\\hspace*{.2in}Supported in part by the DOE
under
grant DE-FG02-92ER40706}\\

Theoretical  Physics Group\\
Martin Fisher School of Physics, Brandeis University\\
Waltham, MA 02454

\end{center}


\renewcommand{\theequation}{1.\arabic{equation}}
\setcounter{equation}{0}

\begin{abstract}
{\normalsize We show that the gluon of  $\cN$=4 Yang--Mills theory
lies on a Regge trajectory, which then implies that the graviton
of $\cN$=8 supergravity also lies on a Regge trajectory.  This is
consistent with the conjecture that  $\cN$=8 supergravity is
ultraviolet finite in perturbation theory.}
\end{abstract}


\section{Introduction}

There has been a great deal of interest in the possibility that
$\cN$=8 supergravity (sugra) has the same degree of divergence as
$\cN$=4 super Yang--Mills theory (YM), and thus is ultraviolet
finite in four-dimensions \cite{bern}.  In particular the $\cN$=8
sugra perturbation expansion is closely related to $\cN$=4 YM
amplitudes \cite{bern,B-B}.  It has been argued by Green, Russo
and Vanhove \cite{green1} that the dualities of $M$-theory imply
that the four-graviton amplitude of $\cN$=8 sugra is ultraviolet
finite in four dimensions. Further they argue \cite{green2} that
even without the duality conjectures, four-dimensional $\cN$=8
sugra might be ultraviolet finite at least up to eight loops.
Clearly the relationship of these two theories is providing new
insights. In this paper we explore the Reggeization program in
both theories.

In our work in collaboration with Grisaru and Tsao
\cite{grisaru1}, and with Grisaru \cite{grisaru2,grisaru3}, we
showed that the elementary fields of renormalizable non-Abelian
gauge theories in four-dimensions lie on Regge trajectories.  In
addition, our old preliminary study \cite{grisaru3} of $\cN$=8
sugra suggested that this {\it might} be true there as well.  In
this paper we extend Ref.\ \cite{grisaru3} to show that the
Reggeization of the gluon in $\cN$=4 YM implies the Reggeization
of the graviton in $\cN$=8 sugra.

The calculations leading to this conclusion involves arcane
technology (to the present generation of theorists).  Therefore we
provide an overview of these methods in Sec.\ 2, which involves
(perhaps unfamiliar) concepts of Mandelstam counting, nonsense
helicity states, etc.  However, since we cannot provide a complete
review of these techniques, the reader should refer to the
original papers, especially ref.\ \cite{grisaru3} for additional
technical details.

In Sec.\ 3 we apply the formulism of Sec.\ 2 to the Reggeization
of the gluon in $N$=4 YM and the graviton in $\cN$=8 sugra.
Concluding remarks are in Sec.\ 4, while several Appendices
collect useful formulae needed in the calculations.  Some of the
information in the appendices is repeated in the text for clarity
of presentation.

\section{Regge Behavior: An Overview}

\renewcommand{\theequation}{2.\arabic{equation}}
\setcounter{equation}{0}

There exist different ways of finding Regge poles in Lagrangian
field theory.  One method consists of summing Feynman diagrams for
large momentum transfer at fixed $s$ in leading logarithm
approximation and recognizing that a Regge trajectory $\a (s)$
corresponds to asymptotic behavior $\sim t^{\a (s)}$.  Another
involves solving analyticity-unitarity integral equations for the
analytic continuation of the scattering amplitude $f(s ,J)$ and
looking for Regge poles directly.  In $\cN$=8 sugra the second
method is applicable, using as input only knowledge of the Born
approximation (as we explain later in this section.)  [Given the
close relationship of $\cN$=4 YM scattering amplitudes to those of
$\cN$=8 sugra, summation of leading logarithms in the latter
theory may also be possible.]  We have used the second method
extensively \cite{grisaru1}--\cite{grisaru3} in the past to find
Regge poles in Yang--Mills theories, which gave the same result as
diagram summation in all cases where a comparison can be made.  We
emphasize that the {\it existence} and {\it number} of Regge
trajectories in a neighborhood of small integral or half-integral
$J$ is {\it independent} of the size of the coupling.  As a
consequence we have control of local properties in $J$ of Regge
trajectories, but not global ones.

One assumes that the (kinematical singularity free) scattering
amplitude $F(s,J)$ can be continued to large $Re \; J$ without
encountering singularities in the angular momentum plane, and that
it can be continued to the left of $Re \; J = N$.  Here one may
encounter singularities such as poles and cuts.  In particular if
one continues $F(s,J)$ to an integer $J=j \leq N$, one may ask if
 \be
F(s,J) \stackrel{?}{=} f_j (s)
 \ee
where the scattering amplitude $f(s,z)$ computed from diagrams has
the form
 \bea
 f_j (s) & = & \frac{1}{2} \int^1_{-1} dz \, P_j (z) f(s,z)\nonumber \\
 & = & \sum^N_{n=0} b_n (s) \; \d_{jn} + {\rm analytic \; in \;} j,
 \eea
where $z = \cos \th$ and $P_j (z)$ is a Legendre polynomial.  [In
renormalizable field theories $N \leq 1$, and if $\cN$=8 sugra is
finite, or at most log.\ divergent, presumably $N \leq 2$.]  The
presence of kronecker delta terms seems to make the equality (2.1)
unlikely, yet one can prove that in certain case that $F(s,j)$ and
$f_j(s)$ {\it must} coincide.  If they do coincide at some value
of $j$, one says that $f_j(s)$ ``Reggeizes" at $J=j$, and that
$F(s,J)$ is analytic in the neighborhood of such $j$.

Mandelstam \cite{mandelstam} has given certain criteria for
establishing whether or not kronecker delta singularities are
present at a given $j$, based on the following counting argument.
Both $f_j(s)$ and $F(s,J)$ have $s$-plane analyticity and
unitarity properties which require that they satisfy certain
$s$-channel dispersion relations, in which the inhomogeneous terms
are the same in both cases, as is the unitarity condition.  Where
the solutions for $f_j(s)$ and $F(s,J)$ may differ is in the value
of possible subtraction constants in the dispersion relations, and
in the positions and residues of CDD poles, which correspond to
singularities not resolved by analyticity and unitarity.  However
both $f_j(s)$ and $F(s,J)$ are subject to identical kinematical
constraints.  If the number of these constraints equals or exceeds
the number of free parameters, the amplitudes {\it must} coincide.
In this case one understands the kronecker delta as the boundary
value of an analytic function, {\it e.g.},
 \be
 \frac{\a (s) - J}{\a (s) - j} \;
 _{\stackrel{\loon}{\scriptstyle \a (s)  \rightarrow J}} \; \d_{Jj} \; .
 \ee
[In field theory the free parameters are masses and coupling
constants.]  Since Mandelstam counting \cite{mandelstam} is
kinematical, it is true to all orders of perturbation theory.
However, the Mandelstam procedure must be carried out for each $j$
separately.

The proof of Mandelstam is delicate as it involves the unitarity
of the theory, which appears to eliminate non-renormalizable
theories, such as massive YM theories without a Higgs mechanism
\cite{dicus}. Nevertheless in this paper we show that the graviton
in $\cN$=8 sugra Reggeizes as a consequence of the Reggeization of
the $\cN$=4 YM massless gluon.

We consider the scattering of massless particles with spin $(p_1 +
p_2 \rightarrow p_3 + p_4)$ as functions of the Mandelstam
variables
 \bea
 s & = & (p_1 + p_2)^2 = 4q^2 \nonumber \\
 t & = & (p_1 - p_3)^2 = - \frac{s}{2} \: (1-z) \\
 u & = & (p_1 - p_4)^2 = - \frac{s}{2} \: (1 + z) \nonumber
 \eea
where $q$ is the center-of-mass momentum, $z = \cos \th$, with
$\th$ the scattering angle.  One considers the integral equation
satisfied by the two-body, kinematical singularity free, helicity
amplitudes
 \bea
\lefteqn{ \tilde{F}_{\l{_3}\l{_4},\l{_1}\l{_2}} (s,J) =
V_{\l{_3}\l{_4},\l{_1}\l{_2}} (s,J)} \nonumber \\
&& + \sum_{\l{_5}\l{_6}} \int \frac{ds^\prime}{s^\prime -s} \:
\rho (s^\prime ) \tilde{F}_{\l{_3}\l{_4},\l{_5}\l{_6}} (s^\prime
,J) \tilde{F}_{\l{_5}\l{_6},\l{_1}\l} (s^\prime ,J)
 \eea
or as matrices
 \be
 \tilde{F}_{\m\l} = V_{\m\l} + \int \: \frac{ds^\prime}{s^\prime
 -s}\tilde{F}_{\m\l{^\prime}} \: \rho_{\l{^\prime}\m{^\prime}}
 \tilde{F}_{\m{^\prime}\l}
 \ee
 where $\l = \l_1 -\l_2, \; \m = \l_3 -\l_4$, and $\rho$ is a
 phase-space factor.  The unitarity condition couples particles of
 different helicity. In the continuation to small $j$ one reaches
 a point where $ j < | \l |, |\m |$ for some value of $\l$ and/or
 $\m$.  The corresponding amplitudes are unphysical, {\it i.e.},
 states $|\l_1 , \l_2 >$ or $|\l_3 , \l_4 >$ with $|\l |$ or $|\m
 |< j$ are ``nonsense" states.

$V_{\m \l} (s,J)$ can be obtained to any given order of
perturbation theory from diagrams.  In particular if we keep only
the lowest order diagrams,
 \be
 V_{\m \l} (s,J) \sim Q_{J-\l{_m}} (z_0 (s)) v_{\m\l} (s)
 \ee
where $\l_m$ = max $(|\l | , |\m |)$ is an integer or
half-integer, and $Q$ is a Legendre function of the 2$^{\rm nd}$
kind, noting that $Q$ functions have poles at negative integers.
Imagine writing (2.6) for sufficiently large $J$ so that all
helicity states are sense, and continuing $J$ to the neighborhood
of some small physical value of $j$ where some of the helicity
states are nonsense.  At such values of $j$ some of the matrix
elements of $V_{\m\l}$ are singular since $Q_{J-\l}$ develops
poles at negative values.  Denote $\tilde{F}_{\m\l}$ by
  \bea
 \left.
  \begin{array}{l}
 \tilde{F}_{ss} \\
 \tilde{F}_{ns} \\
\\
\tilde{F}_{sn}  \\
\tilde{F}_{nn}
\end{array}
\right\}
\begin{array}{c}
\\
{\rm if}\\
\\
\end{array}
\left\{
\begin{array}{l}
|\l | , |\m | \leq j \\
|\l | \leq j,  |\m | > j \\
\\
|\l | > j,  |\m | \leq j \\
|\l | > j,  |\m | > j \; .\end{array} \right.
 \eea
 In the neighborhood of $j$, the
integral equation is of the form
 \bea
 \left( \begin{array}{ll}
 \tilde{F}_{ss} &  \tilde{F}_{sn} \\
 \tilde{F}_{ns} &  \tilde{F}_{nn} \end{array}
 \right)
 =
 \left( \begin{array}{ll}
 -v_{ss} \; \d_{Jj} & v_{sn} (J-j)^{-1/2} \\
 v_{ns} (J-j)^{-1/2} & v_{nn} (J-j)^{-1} \end{array}
 \right) \nonumber \\ [.1in]
+ \int \frac{ds^\prime}{(s^\prime - s)} \: (\tilde{F}) \; \rho \;
(\tilde{F}) \; . \hspace{1.2in}
 \eea
The Born approximation quantities $v_{ss}, \; v_{sn}, \; v_{ns}$,
and $v_{nn}$ are essentially polynomials in $s$.  The solution to
(2.9) is
 \bea
 \tilde{F}_{ss} (s,J) & = & v_{sn{^\prime}} \left[ \frac{K(s)}{J-j
 - v(s) K(s) + {\cal O} (g^4)}\right]_{n{^\prime}n} v_{ns} \nonumber \\
 & + & {\rm regular \; in \;} J
 \eea
 where $K(s)$ is a known integral common to all channels.  Here
 $\tilde{F}_{ss}$ consists of those helicity amplitudes physical
 at $J=j$.

 We find Regge poles with trajectories
  \be
  \a (s) = j + ({\rm eigenvalues \; of}\; v_{nn}) \times K(s) +
  {\cal O} (g^4) \; .
  \ee
The {\it number} of Regge trajectories at $j$ is equal to the {\it
rank} of the nonsense-nonsense matrix $v_{nn}$, with trajectories
given by (2.11).  Eqn.\ (2.11) has the same accuracy as the
leading logarithm approximation in summing diagrams.  Therefore
complete information about the trajectories (but not the residue)
in the neighborhood of $j$ can be obtained by studying $v_{nn}$.
If we keep only contributions of the $t-$ and $u-$channel poles in
Born approximation, this gives the {\it location} of the Regge
poles for values such that $\a (s_o) = j$, with $\a^\prime (s_0)$
correct to $\cO (g^4)$.

Consider (2.10) for $J \rightarrow j$.  Consistency for
Reggeization to occur at $j$ requires the matrix factorization
 \be
 v_{ss} = v_{sn{^\prime}} (v^{-1})_{n{^\prime}n} v_{ns}
 \ee
at $J=j$.  [This should not be confused with the tree
factorization of the Born approximation.]  One has the following
statement \cite{grisaru4}.  A {\it necessary and sufficient}
condition for (2.12) to hold is that the rank of the matrix
 \be
 \left( \begin{array}{ll}
 v_{ss} & v_{sn} \\v_{ns} & v_{nn} \end{array}
 \right) = \mbox{\boldmath $v$}
 \ee
equals that of the nonsense-nonsense matrix $v_{nn}$.  [Recall
that the rank $v_{nn}$ is equal to the number of Regge
trajectories at $j$.]

In the next section we apply the above formalism to gluon-gluon
scattering in $\cN$=4 YM, and graviton-graviton scattering in
$\cN$=8 sugra.  We show that the gluon pole at $j=1$ Reggeizes,
with rank $v_{nn} = 1$ and rank $\mbox{\boldmath $v$} = 1$, and
thus (2.12) is satisfied.  As a consequence, we show that these
relations in $\cN$=4 YM {\it also} imply that the graviton
Reggeizes in $\cN$=8 sugra. That is (2.12) being satisfied for
$\cN$=4 YM also implies that the analogous factorization condition
holds for $\cN$=8 sugra. In both theories the rank $v_{nn} = 1$
and rank  $\mbox{\boldmath $v$} = 1$, though dim $ \mbox{\boldmath
$v$}$ differs.

\section{Helicity Amplitudes}

\renewcommand{\theequation}{3.\arabic{equation}}
\setcounter{equation}{0}

\noindent{\bf A.~ Generalities}

In this section we present the relevant information to justify our
claim of the Reggeization of the graviton and gluon in $\cN$=8
sugra and $\cN$=4 YM respectively.  Since both theories are
supersymmetric, pseudohelicity conservation applies.  For the
scattering amplitude
 $$
 F (\l_3, \l_4 ; \, \l_1,  \l_2 )
  $$
  with helicities $\l_1, \cdots , \; \l_4$ one defines the
  pseudohelicity
 \be
 P(\l_1, \l_2) = \l_1 + \l_2
 \ee
 for the initial state, and similarly for the final state.
 Supersymmetry requires
 \be
 P(\l_1, \l_2) =P( \l_3 , \l_4) \; .
 \ee
That is
 \be
 F(\l_3, \l_4; \l_1,\l_2) =0 \;\;\; {\rm if} \;\;\; \l_1 + \l_2 \neq
  \l_3 + \l_4 \; .
 \ee
As a result, we only need consider $P$=0 states in our discussion
of the Reggeization of the graviton and of the gluon.

In order to extract Regge behavior one must deal with kinematical
singularity free amplitudes, as these are the ones that satisfy
unitarity and analyticity.  Given
 \be
 F(\l_3, \l_4; \l_1,\l_2) = F_{\m\l} (s,t,u)
 \ee
where $\m = \l_3 - \l_4, \; \l = \l_1 - \l_2$, and $s,t,u$ are the
Mandelstam variables,  the kinematical singularity free amplitudes
$\tilde{F}_{\m\l}$ are obtained from
 \bea
 F_{\m\l} (s,t,u) & = & \left[ \sqrt{2} \: \cos
 \frac{\th}{2}\right]^{|\l +\m |}
\left[ \sqrt{2} \: \sin  \frac{\th}{2}\right]^{|\l -\m |}
\tilde{F}_{\m\l} (s,t,u) \nonumber \\[.2in]
& = & (\sqrt{2} )^{|\l + \m | + | \l -\m |} \left[
\sqrt{\frac{-u}{s}} \right]^{|\l +\m |}  \left[
\sqrt{\frac{-t}{s}} \right]^{|\l -\m |} \tilde{F}_{\m\l} (s,t,u)\;
. \eea

A typical singularity free amplitude has the form in Born
approximation
 \be
 \td{F}_{\m\l} = \frac{a}{t} + \frac{b}{u} + \frac{c}{s} \; .
 \ee
Given the kinematical singularity free amplitudes, one obtains the
angular momentum projection
 \be
 F^J_{\l\m} = \frac{1}{2} \int^1_{-1} dz \: C^J_{\l\m} (z)
 \td{F}_{\l\m} (z)
 \ee
where the matrices $C^J_{\l\m}$ we need are tabulated in Appendix
C, with $t = -\frac{s}{2} \, (1-z);$  \linebreak $u= -\frac{s}{2}
$ $(1+z); \; z = \cos \th$. Given the $C^J_{\l\m}$, one computes
the projection (3.7) using
 \be
 \frac{1}{2} \int^1_{-1} \frac{dz \: P_\ell (z)}{a-z} = Q_\ell (a)
 \; .
 \ee
For the continuation to small $j$ use
 \be
 Q_{-\ell} (z) = -\pi (\cot \pi \ell ) P_{\ell-1} (z) + Q_{\ell -1} (z)
 \ee
 where
 \be
 -\pi \cot \pi \ell \; _{\stackrel{{\displaystyle\sim}}{\scriptstyle \ell \rightarrow
 \ell{_0}}} \;
 \frac{1}{(\ell - \ell_0)}
 \ee
 with $\ell$ an integer.

 The $P=0$, kinematic singularity free Born amplitudes for $\cN$=8
 sugra and $\cN$=4  YM are listed in Appendices A and B
 respectively.  Throughout, the explicit factors of $\k$ or $g^2$
 are omitted, though it is obvious how to restore these if needed.
  The amplitudes relevant for the Reggeization of the gluon or
  graviton are the $P$=0, flavor {\it singlet} amplitudes,
  tabulated in (D.1)--(D.3) and (E.1)--(E.6) respectively.  The
  angular momentum projections near $j=1$ and $j=2$ are to be
  found in (D.4)--(D.6) and (E.7)--(E.12) respectively.

  We now turn to providing additional information for the relevant
  $P$=0, two-body amplitudes.

\noindent{\bf B.~ $\cN$=4 YM}

The $P$=0 states that contribute are $|1,-1 >$ and
$|1/2^a,-1/2_b>$; $a$ = 1 to 4, {\it i.e.}, $\l$=2 and 1, with the
former nonsense at $j$=1 and the latter sense, where the helicity
1/2 fermion has ``flavor" $a$.  The kinematical singularity free
amplitudes $\td{F} (1,-1; 1,-1)  ; \; \td{F} (1/2^a,-1/2_b; 1,-1)$
and $\td{F} (1/2^a,-1/2_b; 1/2_c,-1/2^d)$ are to be found in
(B.1)--(B.3), which leads to a 2$\times$2 scattering matrix for
$P$=0.

\noindent{\bf C.~ $\cN$=8 sugra}

As a result of the KLT relations \cite{kawai}, the tree amplitudes
for the 4-point functions of $\cN$=8 sugra can be expressed in
terms of the square of the tree amplitudes of $\cN$=4 YM.  The
relevant $P$=0 $\cN$=8 states are $|2, -2>$, $|3/2^A, -3/2_B>$ and
$|1^{[AB]}, -1_{[CD]}>$, $A$=1 to 8.  Schematically the flavor
singlet, $P$=0 Born amplitudes we need are
 \bea
 \td{M} (2,-2;2,-2) & \sim & \td{F} (1,-1;1,-1)\: \td{F}(1,-1;1,-1)
 \nonumber \\
 \td{M} (2,-2;3/2,-3/2) & \sim & \td{F} (1,-1;1,-1)\: \td{F}(1,-1;1/2,-1/2)
 \nonumber \\
 \td{M} (3/2,-3/2;3/2,-3/2) & \sim & \td{F} (1,-1;1,-1)\: \td{F}(1/2,-1/2;1/2,-1/2)
 \nonumber \\
 \td{M} (2,-2;1,-1) & \sim & \td{F} (1,-1;1/2,-1/2)\: \td{F}(1,-1;1/2,-1/2)
 \nonumber \\
 \td{M} (3/2,-3/2;1,-1) & \sim & \td{F} (1,-1;1/2,-1/2)\: \td{F}(1/2,-1/2;1/2,-1/2)
 \nonumber \\
 \td{M} (1,-1;1,-1) & \sim & \td{F} (1/2,-1/2;1/2,-1/2)\: \td{F}(1/2,-1/2;1/2,-1/2)
 \eea
where the left-hand side are $\cN$=8 sugra amplitudes, and the
right-side are $\cN$=4 YM amplitudes.  Eqn.\ (3.11) implies a
3$\times$3 $P$=0 scattering matrix.  The structure (3.11)
manifests itself in the projections near $J$=2 for the sugra
amplitudes, and its relationship to the projections near $J$=1 for
the YM amplitudes.  This is evident in comparing (D.4)--(D.6) with
(E.7)--(E.11).

Another way of presenting \cite{bern} the KLT relations
\cite{kawai} is
 \bea
 \td{F} & = & - \left[ \frac{C_s}{s} + \frac{C_t}{t} +
 \frac{C_u}{u} \right] \\ [.2in]
 \td{M} & = & - \left[ \frac{C^2_s}{s} + \frac{C^2_t}{t} +
 \frac{C^2_u}{u} \right]
 \eea
which can be confirmed using Appendices A and B.

\noindent{\bf D.~ The helicity matrices}

Generically, for each $j$ the Born approximation is of the form
 \be
 F^J_{\l\m} = \left( \begin{array}{lc}
 v_{ss} \: \d_{Jj} & v_{sn} (J-j)^{-1/2}\\
 v_{ns} (J-j)^{-1/2} & v_{nn} \end{array}
 \right)
 \ee
 which is the first term on the right-hand side of (2.8) given in
 terms of the submatrices $v_{ss}, \; v_{ns} = v_{sn}; \; v_{nn}$.

 For $\cN$=4 YM, we obtain the helicity matrix
 \be
  \mbox{\boldmath $v$}   = \left( \begin{array}{ll}
 v_{nn} & v_{sn} \\
 v_{ns} & v_{ss} \end{array}
 \right)
 \ee
from the $P$=0 states near $j=1$.  Explicit values are in
(D.7)--(D.9), which we repeat for convenience
 \bea
 v_{nn} & = & 4 \nonumber \\
 v_{ns} & = & v_{sn} \; = \; \frac{16}{\sqrt{3}} \nonumber \\
 v_{ss} & = & \frac{64}{3} \; .
 \eea
From (3.15) and (3.16) we have
 \bea
 {\rm rank} \; v_{nn} & = & 1 \nonumber \\
 {\rm rank} \;  \mbox{\boldmath $v$}   & = & 1 \nonumber \\
 v_{ss} & = & v_{sn} (v_{nn})^{-1} \, v_{ns} \; .
 \eea
Eqn. (3.17) combined with Mandelstam counting implies that the
gluon {\it must} Reggeize.

For $\cN$=8 sugra, the analogous helicity matrix for $P$=0 is
[using $V$ to avoid confusion with (3.15)]
 \be
  \mbox{\boldmath $V$}   = \left( \begin{array}{ll}
 V_{nn} & V_{sn} \\
 V_{ns} & V_{ss} \end{array}
 \right)
 \ee
near $j=2$.  Explicit values are in (F.1)--(F.4), with the KLT
relations \cite{kawai} evident in (F.1)--(F.4).

Using (F.1)--(F.4), (3.12) and (3.15)--(3.17) we find that
 \bea
 {\rm rank} \; V_{nn} & = & 1 \nonumber \\
 {\rm rank} \; \mbox{\boldmath $V$}  & = & 1
 \eea
We conclude that the graviton {\it must} Reggeize, as a {\it
consequence} of the Reggeization of the gluon!

\section{Concluding Remarks}

In this paper we have shown that the gluon of $\cN$=4 YM theory
lies on a Regge trajectory.  The factorization condition for this
to hold, (3.17), also implies that the graviton of $\cN$=8 sugra
lies on a Regge trajectory, {\it c.f.} (F.5)--(F.7). In Ref.\
\cite{grisaru3} we only verified that the rank $V_{nn} =1$, but
did not check the factorization condition for $  \mbox{\boldmath
$V$}$, since we presumed that $\cN$=8 sugra was
non-renormalizable.  Here we verify that the factorization
condition for $\cN$=4 YM,  (3.17) then also implies $\det
V_{nn}=1$ and rank $ \mbox{\boldmath $V$}=1$, leading to the
conclusion that the graviton {\it must} Reggeize, which is
consistent with the speculation that $\cN$=8 sugra is ultraviolet
finite. It should be emphasized that this is {\it not} a
holographic result, as both theories are considered in
perturbation theory.

The renormalizability vs.\ non-renormalizability  of a field
theory is not a trivial issue for the Reggeization program, since
at face value there are an infinite number of free parameters for
a non-renormalizable theory, and thus Mandelstam counting would
not apply. An example is massive YM theory without the Higgs
mechanism.  It is known that the factorization condition (2.12)
fails in this case \cite{dicus}, and thus there the gluon does
{\it not} Reggeize. However, $\cN$=8 sugra appears to evade the
difficulties exampled by that massive YM example.

The computation presented in this paper is equivalent to the
leading logarithm approximation in the summation of an infinite
set of diagrams.  It is therefore reasonable to expect that a
leading logarithm summation of diagrams in $\cN$=8
graviton-graviton scattering will reproduce our results.  The
Regge trajectories, computed in perturbation theory, will not lead
to recurrences in weak-coupling, but rather are analogous to the
Regge trajectories of potential scattering.  All orders in
perturbation theory continued to strong-coupling, may well produce
Regge recurrences.

The other fundamental fields of the Lagrangians of $\cN$=8 sugra
and $\cN$=4 YM should Reggeize as well, as a consequence of the
unbroken SU(8) and SU(4) flavor symmetries (respectively) of the
theories.

Further exploration of the possible implications of an ultraviolet
finite $\cN$=8 sugra promises to be a fruitful enterprize. It is
an important issue of principle to know whether or not $\cN$=8
sugra is a finite quantum theory of gravity {\it distinct} from
string theory.  Given that $\cN$=8 sugra contains Regge poles, we
speculate that is is not distinct.

\noindent{\bf Acknowledgements}

We thank Albion Lawrence for stimulating conversations.  We are
also appreciative of our old collaboration with Marc Grisaru on
this subject.

\pagebreak

\renewcommand{\theequation}{A.\arabic{equation}}
\setcounter{equation}{0}

\noindent{\large\bf Appendix A}

\noindent{\bf $\cN$=8 sugra: Kinematical Free Amplitudes: $P=0$}

\vspace*{-.3in}

\bea
 \tilde{M} (2, -2; 2, -2 ) & = & \frac{-s^2}{16} \: \left[
 \frac{1}{t} + \frac{1}{u} \right] \; , \\[.1in]
\tilde{M} \left( \frac{-3}{2}^A , \frac{-3}{2}_B ; 2, -2 \right) &
= & - \tilde{M} \left( 2, -2; \frac{3}{2}_B, \frac{-3}{2}^A
\right)
\nonumber \\[.1in]
& = & \frac{-\d^A_Bs^2}{16} \: \left[ \frac{1}{u} +
\frac{1}{t}\right] \; , \\ [.1in]
 \tilde{M} \left( \frac{3}{2}^A ,
\frac{-3}{2}_B ; \frac{3}{2}_C , \frac{-3}{2}^D \right) & = &
\frac{1}{8} \left(\frac{s^3}{u} \right) \left[
\frac{\d^A_B\d^D_C}{s} + \frac{\d^A_C\d^D_B}{t} \right] \;
,\\[.1in]
\tilde{M} (1^{AD}, - 1_{BC} ; 2, -2) & = & \tilde{M}(2, -2;
1_{AD}, -1^{BC})^* \nonumber \\[.1in]
& = & \frac{-s^2}{16} \left(\frac{1}{t} + \frac{1}{u} \right)
\d^{AD}_{BC} \; , \\[.1in]
 \tilde{M} \left( \frac{3}{2}^A , \frac{-3}{2}_B ; 1_{CF}, -1^{DE}
 \right) & = & - \tilde{M} \left( 1^{CF}, -1_{DE}; \frac{3}{2}_A,
 \frac{-3}{2}^B \right)^* \nonumber \\ [.1in]
 & = & \frac{1}{8} \left(\frac{s^3}{u} \right) \left\{ \frac{1}{t}
 \;
 [\d^A_F\d^{DE}_{BC} -\d^A_C\d^{DE}_{BF} ] - \frac{1}{s} \;
 \d^A_B\d^{DE}_{CF} \right\} \; ,
  \eea
where throughout (*) is an SU(8) conjugation which raises and
lowers indices {\it only}:
 \bea
 \tilde{M}(1^{AH}, - 1_{BC}; 1_{DE}, -1^{FG}) & = & \frac{-s^2}{4}
 \: \left\{ \frac{1}{u} \frac{1}{4!} \;
 \ve^{AHFGMNOP}\ve_{BCDEMNOP}\right.
 \nonumber \\ [.1in]
 && + \; \left. \frac{1}{s} \; \d^{AH}_{BC}\d^{FG}_{DE} + \frac{1}{t} \;
 \d^{AH}_{DE}\d^{FG}_{BC} \right\}
\eea
 with
 \vspace{-.3in}
 \bea
 \d^{AB}_{CD} & = & \d^A_C\d^B_D - \d^A_D\d^B_C \; , \\
 \d^{ABC}_{DEF} & = & \d^A_D\d^{BC}_{EF} + \d^B_D\d^{CA}_{EF} +
 \d^C_D\d^{AB}_{EF} \nonumber \\
 & = & \frac{1}{5!} \; \ve^{ABCGHKLM} \ve_{DEFGHKLM} \; , \\
 \d^{ABDDE}_{FGHKL} & = & \frac{1}{3!} \; \ve^{ABCDEMNO} \ve_{FGHKLMNO}
 \eea
where the flavor indices are $A =$ 1 to 8.

\pagebreak

\noindent{\large\bf Appendix B}

\renewcommand{\theequation}{B.\arabic{equation}}
\setcounter{equation}{0}

\noindent{\bf $\cN$=4 YM: Singularity-Free Amplitudes}

Pseudohelicity (0)

\bea
 \td{F}(1, -1; 1, -1) & = & s \left[ \frac{\a}{t} + \frac{\b}{u}
 \right] \\ [.1in]
 \tilde{F} \left( \frac{1}{2}^a, - \frac{1}{2}_b ; 1, -1 \right) & = &
 \d^a_b \; s \left[ \frac{\a}{t} + \frac{\b}{u}
 \right] \\[.1in]
 \tilde{F} \left( \frac{1}{2}^a, - \frac{1}{2}_b ; \frac{1}{2}_c ,
\frac{1}{2}^d \right) & = & 2 \left[ s \: \d^a_c \d^d_b + t \:
\d^a_b \d^d_c \right] \left[ \frac{\a}{t} + \frac{\b}{u} \right]
\eea
 where
 \bea
 \a & = & f_{ikn}f_{nj\ell} \nonumber \\
 \b & = & f_{i\ell n}f_{njk}
 \eea
 are products of the structure constants of the gauge group, and
 the flavor indices are $a =$ 1 to 4.

\newpage

\noindent{\large\bf Appendix C}
\renewcommand{\theequation}{C.\arabic{equation}}
\setcounter{equation}{0}
 {\bf The functions $C^J_{\l\m}$}

 \bea
 C^J_{44} & = &
 \frac{(J+1)(J+2)(J+3)(J+4)}{(2J-5)(2J-3)(2J-2)(2J+1)} \; P_{J-4}
 \\ [.2in]
 & &+
 \frac{4(J+2)(J+3)(J+4)}{(2J-3)(2J-1)(2J+1)} \, P_{J-3}  +
 \frac{28(J-3)(J+2)(J+3)(J+4)}{(2J-5)(2J-1)(2J+1)(2J+3)} \,
 P_{J-2}+ \cdots , \nonumber \\ [.3in]
C^J_{33} & = &
 \frac{(J+1)(J+2)(J+3)}{(2J-3)(2J-1)(2J+1)} \; P_{J-3}
 \nonumber \\[.2in]
 && +
 \frac{3(J+2)(J+3)}{(2J-1)(2J+1)} \; P_{J-2}  +
 \frac{15(J-2)(J+2)(J+3)}{(2J-3)(2J+1)(2J+3)} \;
 P_{J-1}+ \ldots \; ,\\[.3in]
 C^J_{34} & = & \sqrt{(J+4)(J-3)} \; \left\{
\frac{(J+1)(J+2)(J+3)}{(2J-5)(2J-3)(2J-1)(2J+1)} \; P_{J-4}
\right.
\nonumber \\[.2in]
&&  + \frac{3(J+2)(J+3)}{(2J-3)(2J-1)(2J+1)} \; P_{J-3} \nonumber
\\[.2in]
&&  + \left. \frac{7(J+2)(J+3)(2J-7)}{(2J-5)(2J-1)(2J+1)(2J+3)} \;
P_{J-2} + \cdots \right\} \; ,\\[.3in]
 C^J_{24} & = & \sqrt{(J-2)(J-3)(J+3)(J+4)} \; \left\{
\frac{(J+1)(J+2)}{(2J-5)(2J-3)(2J-1)(2J+1)} \; P_{J-4} \right.
\nonumber \\[.2in]
&&  + \frac{2(J+2)}{(2J-3)(2J-1)(2J+1)} \; P_{J-3} \nonumber
\\[.3in]
&&  + \left. \frac{4(J-6)(J+2)}{(2J-5)(2J-1)(2J+1)(2J+3)} \;
P_{J-2} + \cdots \right\} \; ,\\[.3in]
 C^J_{23} & = & \sqrt{(J-2)(J+3)} \; \left\{
\frac{(J+1)(J+2)}{(2J-3)(2J-1)(2J+1)} \; P_{J-3}
  + \frac{2(J+2)}{(2J-1)(2J+1)} \; P_{J-2}
\right.\nonumber
\\[.2in]
&&  + \left. \frac{5(J-3)(J+2)}{(2J-3)(2J+1)(2J+3)} \;
P_{J-1} + \cdots \right\} \; ,
\\[.3in]
 C^J_{22} & = &
 \frac{(J-1)J}{(2J+1)(2J+3)} \; P_{J+4} + \frac{2(J-1)}{(2J+1)}
  \; P_{J+1} \nonumber
 \\ [.2in]
&& +
 \frac{6(J-1)(J+2)}{(2J-1)(2J+3)} \; P_J + \frac{2(J+2)}{(2J+1)}
  \; P_{J-1} \nonumber
 \\ [.2in]
&& +
 \frac{(J+1)(J+2)}{(2J-1)(2J+1)} \; P_{J-2}  + \cdots
  \; ,
\\ [.3in]
 C^J_{11} & = &
 \frac{J}{(2J+1)} \; P_{J+1} + P_J + \frac{(J+1)}{(2J+1)}
  \; P_{J-1}  + \cdots \; ,
 \\ [.3in]
 C^J_{12} & = & \sqrt{(J-1)(J+2)} \; \left\{
\frac{-J}{ (2J+1)(2J+3)} \; P_{J+2} - \frac{1}{(2J+1)}
 \; P_{J+1} \right.
\nonumber \\[.2in]
&&  \frac{-3}{(2J-1)(2J+3)} \; P_J + \frac{1}{(2J+1)} P_{J-1}
\nonumber
\\[.2in]
&&  + \left. \frac{(J+1)}{(2J-1)(2J+1)} \; P_{J-2} + \cdots
\right\} \; .
  \eea

\newpage

\noindent{\large\bf Appendix D}

\renewcommand{\theequation}{D.\arabic{equation}}
\setcounter{equation}{0}


 \noindent{\bf a)~ $\cN$=4 flavor singlet amplitudes}

 \bea
 \tilde{F} (1,-1;1,-1) & = & s \left[ \frac{\a}{t} + \frac{\b}{u}
 \right] \\ [.1in]
 \tilde{F} (1/2,-1/2;1,-1) & = & 4s \left[ \frac{\a}{t} + \frac{\b}{u}
 \right] \\ [.1in]
\tilde{F} (1/2,-1/2;1/2,-1/2) & = & 8 [s + 4t] \left[ \frac{\a}{t}
+ \frac{\b}{u}
 \right]
\eea with $\a ,  \b$ in (B.4).

\vspace*{.2in}

 \noindent {\bf b)~ Angular momentum projections
near $J$=1}

 Using (2.9) and Appendix C
 \bea
 F^J_{22} (1,-1;1,-1) & = &  \frac{v_{nn} (\a - \b )}{(J-1)}  \\ [.1in]
 F^J_{12} (1/2,-1/2;1,-1) & = &  \frac{v_{sn} (\a - \b )}{\sqrt{J-1}}  \\ [.1in]
 F^J_{11} (1/2,-1/2;1/2,-1/2) & = & v_{ss} (\a -\b ) \d_{J1}
\eea where $(\a - \b )$ belongs to the adjoint representation of
the group, and
 \bea
 v_{nn} & = & 4 \\
 v_{sn} & = &  v_{ns} \; = \; \frac{16}{\sqrt{3}} \\
 v_{ss} & = & \frac{64}{3} \; .
 \eea
Note that (2.11) is satisfied, and thus
 \be
 \det  \mbox{\boldmath $v$} = 0
 \ee
 where the helicity matrix $\mbox{\boldmath $v$}$ is defined in
 (2.13).

\newpage

\noindent{\large\bf Appendix E}
\renewcommand{\theequation}{E.\arabic{equation}}
\setcounter{equation}{0}

\noindent{\bf a)~ $\cN$=8 flavor singlet amplitudes}

 \vspace*{-.3in}

 \bea
 \tilde{M} (2,-2;2,-2) & = & - \frac{s^2}{16} \left[ \frac{1}{t} + \frac{1}{u}
 \right] \\ [.1in]
 \tilde{M} (2,-2;3/2,-3/2) & = & - \frac{-s^2}{2} \left[ \frac{1}{t} + \frac{1}{u}
 \right] \\ [.1in]
 \tilde{M} (3/2,-3/2;3/2,-3/2) & = &  2 \left( \frac{s^3}{u}\right) \left[ \frac{8}{s} +
 \frac{1}{t}
 \right] \\ [.1in]
 \tilde{M} (2,-2;1,-1) & = & - \frac{7s^2}{2} \left[ \frac{1}{t} + \frac{1}{u}
 \right] \\ [.1in]
\tilde{M} (3/2,-3/2;1,-1) & = &  \frac{7s^3}{u} \left[ \frac{1}{t}
+ \frac{4}{u}
 \right] \\ [.1in]
 \tilde{M} (1,-1;1,-1) & = & - 28 s^2  \left[ \frac{15}{u} + \frac{28}{s}
 + \frac{1}{t} \right] \; .
 \eea

 \noindent{\bf b)~ Angular momentum projections near $J$=2}

 The structure of (E.7)--(E.14) below reflects that $\cN$=8 tree
 amplitudes can be expressed in terms of the squares of $\cN$=4 YM
 tree amplitudes.
 \bea
 M^J_{44} (2,-2;2,-2) & = & \frac{3s}{8} \;
 \frac{v_{nn}v_{nn}}{(J-2)} \\[.1in]
 M^J_{43} (2,-2;3/2,-3/2) & = & \frac{-3_i\sqrt{2} \, s}{8} \;
 \frac{v_{nn}v_{ns}}{(J-2)} \\[.1in]
 M^J_{33} (3/2,-3/2;3/2,-3/2) & = & \frac{-3s}{4} \;
 \frac{v_{nn}v_{ss}}{(J-2)} \\[.1in]
M^J_{42} (2, -2;1,-1) & = & \frac{-21is}{32} \; \sqrt{\frac{6}{5}}
\; \frac{(v_{ns})(v_{ns})}{\sqrt{J-2}} \\[.1in]
M^J_{32} (3/2, -3/2;1,-1)
& = & \frac{-21s}{16} \; \sqrt{\frac{3}{5}} \;
\frac{(v_{ns})(v_{ss})}{\sqrt{J-2}} \\[.1in]
M^J_{22}  (1,-1;1,-1)  & = & \frac{-(21)^2s}{320} \; (v_{ss})
(v_{ss}) \d_{J2}
 \eea

\newpage

\noindent{\large\bf Appendix F}
\renewcommand{\theequation}{F.\arabic{equation}}
\setcounter{equation}{0}

\noindent{\bf  $\cN$=8 helicity matrices}

Equations (E.7)--(E.12) can be written as in (2.8), but using
$V_{ss}, \; V_{sn}$, and $V_{nn}$ to distinguish these from the
analogous matrices $v_{ss}, \; v_{sn}$, and $v_{nn}$ of $\cN$=4
YM. [{\it c.f.} Appendix D.]
 \be
 V_{nn} = \frac{3s}{8} \: v_{nn} \left[ \begin{array}{lr}
 v_{nn} & -i \sqrt{2} \, v_{ns}\\
-i \sqrt{2} \, v_{ns} & -2 \, v_{ss}
\end{array} \right]
\ee
 from (E.7)--(E.9)
 \be
 V_{sn} = \frac{-21}{32} \: \sqrt{\frac{3}{5}} \; s \:
  v_{ns} \left[ \begin{array}{cc}
i \sqrt{2} & v_{ns}\\
 ~~~~~2 \; v_{ss}\end{array}
 \right]
\ee from (E.10)--(E.11).
 \be
 V_{ss} = \frac{-(21)^2}{320} \; s \: v_{ss} [v_{ss}]
 \ee
 from (E.12).  The 3$\times$3 matrix  $\mbox{\boldmath$V$}$ is\\
\be
  \mbox{\boldmath $V$} =
  \left[
 \begin{array}{l|l}
  V_{nn} & V_{sn}\\
  \hline
  V_{ns} & V_{ss}
 \end{array}
 \right] \; .
 \ee
One has
 \bea
 \det \; V_{nn} & = & 0 \;\;\;\;;  \nonumber \\
 {\rm rank} \; V_{nn} & = & 1
 \eea
 as a consequence of the factorization condition (2.12) for
 $\cN$=4 YM, {\it i.e.},
  \be
  v_{nn} v_{ss} - v_{ns} v_{sn} = 0 \; .
  \ee
Thus there is only one flavor singlet Regge trajectory at $J$=2 in
$\cN$=8 sugra.

From (F.1)--(F.4) and (F.6) one finds
 \be
{\rm rank} \; \mbox{\boldmath $V$} = 1
 \ee
 which implies that the graviton in $\cN$=8 sugra Reggeizes as a
 consequence of the Reggeization of the gluon in $\cN$=4 YM.

\end{document}